\documentclass[twocolumn, astrosymb]{aastex631}

\usepackage{amssymb, amsmath}
\usepackage{CJKutf8}
\usepackage{bm}
\usepackage{hyperref}
\usepackage{fontawesome5}
\usepackage{academicons}
\usepackage{graphicx}
\usepackage{float}
\shorttitle{Stratified fossil record of halo mass history}
\shortauthors{Xu et al.}

\graphicspath{{./}{figures/}}
\hypersetup{colorlinks=true,
            citecolor=MidnightBlue,
            linkcolor=MidnightBlue,
            filecolor=magenta,
            urlcolor=MidnightBlue}
\defcitealias{Huang2018b}{Paper~I}	
\defcitealias{Huang2018c}{Paper~II}	

\def\apjs{{ApJS}}

\def\h{\hskip -3 mm}

\def\deg{$^{\circ}$}

\def\lax{{$\mathrel{\hbox{\rlap{\hbox{\lower4pt\hbox{${\sim}$}}}\hbox{$<$}}}$}}
\def\gax{{$\mathrel{\hbox{\rlap{\hbox{\lower4pt\hbox{${\sim}$}}}\hbox{$>$}}}$}}
\def\simlt{\lower.5ex\hbox{$\; \buildrel < \over {\sim} \;$}}
\def\simgt{\lower.5ex\hbox{$\; \buildrel > \over {\sim} \;$}}

\def\logm10{{$\log (M_{\star,10\ \mathrm{kpc}}/M_{\odot})$}}
\def\logm30{{$\log (M_{\star,30\ \mathrm{kpc}}/M_{\odot})$}}
\def\logm50{{$\log (M_{\star,50\ \mathrm{kpc}}/M_{\odot})$}}
\def\logm100{{$\log (M_{\star,100\ \mathrm{kpc}}/M_{\odot})$}}

\def\mhalo{{$M_{\rm halo}$}}

\def\mh200b{{$M_{\mathrm{200b}}$}}
\def\mh200c{{$M_{\mathrm{200c}}$}}

\def\insitu{{\textit{in situ}}}
\def\exsitu{{\textit{ex situ}}}

\def\mdpl2{\texttt{MDPL2}}

\def\tng{{\tt IllustrisTNG}}

\definecolor{LightGray}{gray}{0.85}
\definecolor{Tab1}{RGB}{114, 158, 206}
\definecolor{Tab2}{RGB}{255, 158,  74}
\definecolor{Tab3}{RGB}{103, 191,  92}
\definecolor{Tab4}{RGB}{174, 199, 232}
\definecolor{Tab5}{RGB}{255, 187, 120}
\definecolor{Tab6}{RGB}{152, 223, 138}
\definecolor{Tab7}{RGB}{255, 152, 150}
\definecolor{Tab8}{RGB}{197, 176, 213}
\definecolor{hpurple}{HTML}{7E16DF}

\newcommand{\code}[1]{\textbf{\texttt{#1}}} % Code or software name
\defcitealias{Xu2026arXiv}{Paper~I}

\begin{document}
\begin{CJK*}{UTF8}{gbsn}

\title{The Connection between Halo Assembly History and the Stellar Distribution II: A Radially Stratified Fossil Record of Past Halo Mass}

\correspondingauthor{Song Huang}
\email{shuang@tsinghua.edu.cn, xus21@mails.tsinghua.edu.cn}

\author[0000-0002-4460-0409]{Shuo Xu (许朔)}
\affiliation{Department of Astronomy, Tsinghua University, Beijing 100084, China}

\author[0000-0003-1385-7591]{Song Huang (黄崧)}
\affiliation{Department of Astronomy, Tsinghua University, Beijing 100084, China}

\author[0000-0002-2897-6326]{Conghao Zhou (周丛浩)}
\affiliation{Physics Department, University of California, Santa Cruz, CA 95064, USA}
\affiliation{Santa Cruz Institute for Particle Physics, Santa Cruz, CA 95064, USA}

\author[0000-0002-3677-3617]{Alexie Leauthaud}
\affiliation{Department of Astronomy and Astrophysics, UCO/Lick Observatory, University of California, 1156 High Street, Santa Cruz, CA 95064, USA}

\author[0000-0001-9568-7287]{Benedikt Diemer}
\affiliation{Department of Astronomy, University of Maryland, College Park, MD 20742, USA}

% -------------------------------------------------------------------------------------------- %
% Abstract & Keywords
% -------------------------------------------------------------------------------------------- %
\begin{abstract}

    The spatial stellar distribution in massive central galaxies preserves a stratified fossil record of their host halos' assembly history. Using a halo-mass-complete sample from the {\tt IllustrisTNG-300} hydrodynamical simulation at $z=0.4$, we quantify this connection with two diagnostics. The fossil correlation $\tilde{\rho}(z)$ measures how strongly a present-day galaxy property retains information about the progenitor halo mass at each redshift. The rank retention fraction $f_{\rm ret}(z)$ measures how many halos selected by that property today were among the most massive halos at each epoch. The central stellar mass at different galactocentric radii exhibits non-monotonic fossil correlations with past halo mass, peaking at intermediate redshifts that systematically shift from $z\sim1$ in the inner regions to $z\sim0.7$ in the outer stellar halos. These peak redshifts are robust from group to cluster scales. Partial correlations with halo accretion rate show that each peak marks the end of the epoch when that component's buildup was sensitive to halo accretion. The rank retention fraction shows that the annular stellar masses of the central galaxy beyond 10 kpc are less biased tracers of the most massive halos at intermediate redshifts than richness. This stratified fossil record originates mainly from the \exsitu{} phase of galaxy assembly. The outer stellar halo also traces the satellite population at intermediate redshift better than present-day richness. These results establish that the stellar profile of massive central galaxies encodes a radially stratified fossil record of halo assembly history.
    
\end{abstract}
% The AAS Journals now uses Unified Astronomy Thesaurus concepts:
% https://astrothesaurus.org

% -------------------------------------------------------------------------------------------- %
% Introduction
% -------------------------------------------------------------------------------------------- %

\section{Introduction} 
    \label{sec:intro}

    Massive dark matter halos ($M_{\rm 200c}\gtrsim10^{13}\,M_\odot$) host the most massive galaxies in the present-day Universe, and their abundance and clustering are powerful cosmological probes (e.g., \citealt{Allen2011ARA&A}; \citealt{Kravtsov2012ARA}). At fixed halo mass, however, halos with different assembly histories host different galaxy populations and cluster differently, an effect known as halo assembly bias (e.g., \citealt{Gao2005MNRAS}; \citealt{Wechsler2006ApJ}; \citealt{Gao2007MNRAS}). The galaxy--halo connection is therefore inseparable from halo assembly history, and recovering that history from observable galaxy properties is central to both the physics of massive-galaxy formation and unbiased cosmological inference (\citealt{Wechsler2018ARA&A}; \citealt{Behroozi2019MNRAS}).

    Halo assembly history shapes our measurements in two distinct ways. At fixed present-day mass, secondary halo properties imprint on galaxy observables and on the lensing and clustering signals (assembly bias). Across cosmic time, halo-mass rankings are not conserved, so a sample selected at low redshift does not faithfully trace the most massive halos at earlier epochs (progenitor bias; e.g., \citealt{vanDokkum2001ApJ}; \citealt{Leja2013ApJ}; \citealt{Behroozi2013ApJL}; \citealt{Torrey2015MNRAS}). Both effects motivate the search for observables that encode a halo's \emph{time-resolved} assembly history.

    Massive halos provide two stellar tracers that naturally divide this history in time: the central galaxy, assembled largely through early accretion, and the satellite population, continually replenished by recent accretion (e.g., \citealt{Golden-Marx2018ApJ}; \citealt{Bradshaw2020MNRAS}; \citealt{Wang2024MNRAS}; \citealt{Leidig2025arXiv}; \citealt{Xu2026arXiv}). Because the two tracers are sensitive to different epochs, selecting halos by one or the other should preferentially pick halos with different assembly histories, even at the same halo mass. In \citet[][hereafter \citetalias{Xu2026arXiv}]{Xu2026arXiv}, we confirmed this directly: at fixed halo mass, halos selected by the central stellar mass assembled systematically earlier than those selected by the satellite richness.
    
    That analysis, however, compressed the central galaxy into essentially a single stellar-mass measurement. The stars of a central galaxy span a wide range of radii, and their spatial arrangement carries more information than any single measurement. In the two-phase, inside-out picture of massive galaxy formation (e.g., \citealt{Naab2009ApJl}; \citealt{Oser2010ApJ}; \citealt{vanDokkum2010ApJ}; \citealt{Huang2013ApJL}), a compact core forms early from \insitu{} star formation, while an extended envelope is accreted later from disrupted and stripped satellites (e.g., \citealt{RodriguezGomez2016MNRAS}; \citealt{Cannarozzo2023MNRAS}). Because \exsitu{} accretion drives substantial size growth, successively larger radii should be assembled at successively later epochs. The same principle is well established for the dark matter itself: the density profile of a halo retains a memory of its assembly history \citep{Ludlow2013MNRAS}. The stellar profile offers an observable counterpart of this record, though a distinct one: the accreted stars are contributed by the more strongly bound material, the disrupted satellite galaxies \citep{Deason2021MNRAS, Manuwal2025MNRAS}. The stellar profile should therefore preserve a stratified fossil record of the host halo's assembly history.
    
    This record is not guaranteed to survive. The relative contributions of \insitu{} star formation and accretion vary from halo to halo. The \insitu{} component reflects each halo's star formation history, including late rejuvenation triggered by gas-rich mergers. The accreted envelope is shaped by the merger history: major mergers redistribute mass throughout the profile, while minor mergers mainly build the outer stellar halo (e.g., \citealt{Hilz2012MNRAS, Hilz2013MNRAS}; \citealt{Cooper2013MNRAS}; \citealt{Karademir2019MNRAS}). Later accretion can also cover up the imprints of earlier epochs. Whether the present-day profile still preserves the expected ordering is an empirical question with three parts: how much of the assembly history the stellar profile remembers, which radii encode which epochs, and how long this memory survives later growth. Answering them extends the stellar--halo mass relation beyond a single proxy and the two-observable plane of \citetalias{Xu2026arXiv}, toward a temporally resolved mapping.

    To address these questions quantitatively, we introduce two diagnostics: the \emph{fossil correlation}, which measures how strongly a present-day observable preserves information about the halo's mass at each past epoch, and the \emph{rank retention fraction}, which measures how faithfully a sample selected by that observable traces the most massive halos back through cosmic time. We define both in Section~\ref{ssec:assess} and measure them in the {\tt TNG300} simulation. By demonstrating that the stellar profile carries time-ordered, radially separated assembly information, our results motivate extending the stellar--halo mass relation into a forward model that maps a halo's mass accretion history and initial conditions onto the stellar mass profile and satellite population of its central galaxy, a direction anticipated by \citet{Huang2020MNRAS} and \citetalias{Xu2026arXiv}.

    The \tng{} project adopts a $\Lambda$CDM cosmology with $\Omega_m=0.3089$,~$\Omega_{\Lambda}=0.6911$,~$\Omega_b=0.0486$, and~$H_0=67.74$~km~s$^{-1}$~Mpc$^{-1}$.
    Throughout this paper, the halo mass $M_{\rm halo}$ refers to $M_{\rm 200c}$, the mass enclosed within a sphere whose mean density is 200 times the critical density of the Universe.

% -------------------------------------------------------------------------------------------- %
% Data
% -------------------------------------------------------------------------------------------- %

\section{Simulation Data} 
    \label{sec:data}
    
    Following \citetalias{Xu2026arXiv}, we use an \mhalo{}-complete sample of massive halos from {\tt TNG300-1}, the largest-volume hydrodynamical simulation of the \tng{} project (\citealt{Pillepich2018MNRAS}; \citealt{Springel2018MNRAS}; \citealt{Nelson2018MNRAS}; \citealt{Naiman2018MNRAS}; \citealt{Marinacci2018MNRAS}), whose data are publicly available \citep{Nelson2019ComAC}\footnote{\url{https://www.tng-project.org/}}. This sample contains 3,388 halos with $\log[M_{\rm 200c}/M_\odot]>13$ at $z=0.4$. {\tt TNG300-1} has a baryonic mass resolution of $1.1\times10^{7}\,M_\odot$ and a softening length of 1.48~kpc, well below the smallest aperture used here (10~kpc). We choose $z=0.4$ to match the mean redshift of the massive galaxies observed in deep Hyper Suprime-Cam imaging \citep[e.g.,][]{Huang2018}.

    For each halo's central galaxy, we project its stellar particles in the central region of its friends-of-friends (FoF) group, excluding those bound to satellite galaxies, along the simulation box $ z$-axis onto a $300\times300$-pixel map with 1~kpc pixels, yielding a 2D stellar mass map. From these maps, we measure the central stellar mass within elliptical apertures that follow the average isophotal shape of each galaxy, as detailed in \citet{Xu2025ApJ} and \citetalias{Xu2026arXiv}.

    To measure richness and its evolution, we use the positions and stellar masses of all subhalos from $z=3$ to $z=0.4$, and we follow the progenitors of each halo with the SubLink merger trees (\S\ref{ssec:hist}).

% -------------------------------------------------------------------------------------------- %
% Methods
% -------------------------------------------------------------------------------------------- %

\section{Methods} 
    \label{sec:methods}

    We measure two observables for each halo: the stellar mass of the central galaxy and the satellite richness. Both quantities serve as halo mass proxies with direct observational counterparts in deep imaging surveys, although the richness used here is counted in 3D within $R_{\rm 200c}$ and therefore excludes projection effects.

\begin{figure*}
    \centering
    \includegraphics[width=1\linewidth, trim=1cm 0 0 0]{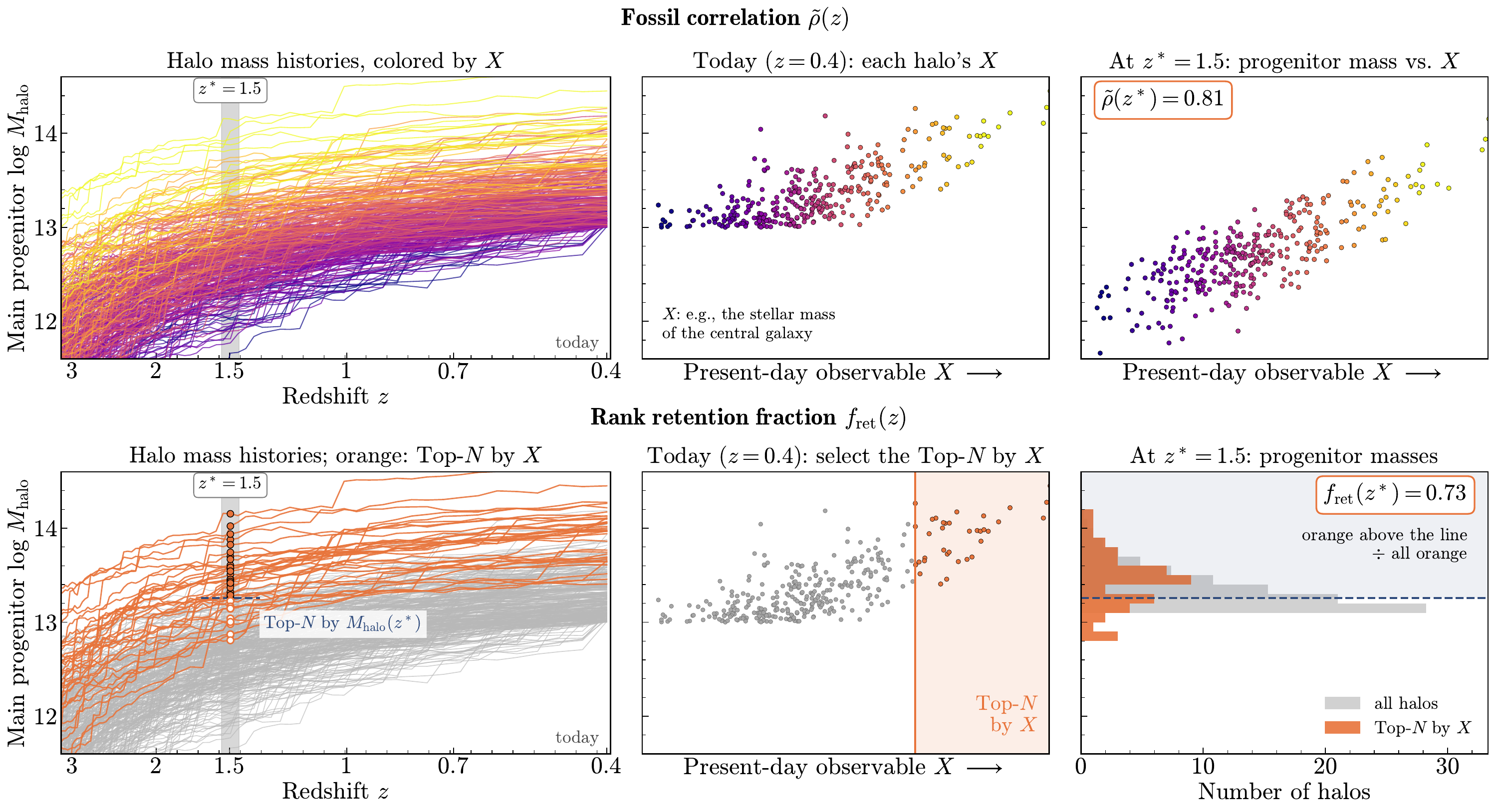}
    \caption{
        Schematic illustration using mock data of the two diagnostics used in this work.
        \textit{Top row: the fossil correlation $\tilde{\rho}(z)$:} the correlation, across halos, between a present-day observable $X$ and the mass of each halo's main progenitor at an earlier epoch $z$.
        \textbf{Left}: main-progenitor mass histories of halos selected today ($\log M_{\rm halo} > 13$ at $z = 0.4$), colored by a present-day observable $X$ (e.g., the stellar mass of the central galaxy).
        \textbf{Middle}: $X$ versus present-day halo mass.
        \textbf{Right}: $X$ versus the main-progenitor mass at an earlier epoch $z^* = 1.5$.
        The correlation of this cloud defines $\tilde{\rho}(z^*)$.
        \textit{Bottom row: the rank retention fraction $f_{\rm ret}(z)$:} the fraction of the Top-$N$ halos selected by $X$ today whose main progenitors are among the Top-$N$ most massive halos at $z$.
        \textbf{Middle}: the Top-$N$ halos by $X$ are selected today (orange).
        \textbf{Left}: their mass histories. Circles mark their progenitor masses at $z^*$ and the dashed segment marks the Top-$N$ mass threshold at that epoch. Filled circles lie above the threshold, open circles below.
        \textbf{Right}: progenitor masses at $z^*$ for the selected halos (orange) and for the full halo population (gray, from the halo mass function).
        $f_{\rm ret}(z^*)$ is the fraction of the orange sample above the dashed line.}
    \label{fig:illustration}
\end{figure*}

% -------------------------------------------------------------------------------------------- %
\subsection{Central Stellar Mass Measurement}
    \label{ssec:Mstarmethod}

    Central galaxies in massive halos typically have smooth, elliptical stellar distributions. Following \citet{Huang2018} and \citet{Li2022}, we fit 1D elliptical isophotes to each 2D stellar mass map with the \texttt{isophote} module of \texttt{Photutils} (\citealt{Photutils}; based on the algorithm of \citealt{Jedrzejewski1987MNRAS}; see \citealt{Xu2025ApJ} for details), and we use them to define the average ellipticity and position angle of each galaxy.

    We then measure stellar masses within elliptical apertures of fixed physical size that follow this average isophotal shape. We use five apertures with radii of 10, 30, 50, 100, and 150~kpc, denoted $M_{\star,10}$, $M_{\star,30}$, and so on. We also compute the stellar mass in outer annuli as differences between aperture masses (e.g., $M_{\star,[50,100]} \equiv M_{\star,100} - M_{\star,50}$), which isolates the diffuse stellar halo within a given radial range.

% -------------------------------------------------------------------------------------------- %
\subsection{Richness Measurement}
    \label{ssec:richmethod}

    Richness, the number of galaxies above a stellar mass threshold within a specified region around a halo, is an established halo mass proxy used by cluster finders such as redMaPPer \citep{Rykoff2014ApJ} and CAMIRA \citep{Oguri2014MNRAS}. Following \citetalias{Xu2026arXiv}, we measure it from subhalo catalogs at both the present day and earlier epochs.

    For the present-day richness, we adopt the ``true'' richness of \citetalias{Xu2026arXiv}: the number of satellite galaxies with stellar mass $M_{\star} > 10^{10}~M_{\odot}$ within a spherical region centered on the central galaxy with radius $R_{\rm 200c}$ ($\lambda_{10, R_{200}}$). For the historic richness, we use the halo and subhalo catalogs at earlier redshifts. To better count all resolved satellites, we apply a lower stellar mass threshold of $M_{\star} > 10^{9}~M_{\odot}$ and count satellites within the $R_{\rm 200c}$ sphere at each corresponding epoch ($\lambda_{9, R_{200}}$). 

% -------------------------------------------------------------------------------------------- %
\subsection{Mass Accretion History and Accretion Rate}
    \label{ssec:hist}

    We characterize the assembly history of each halo with the SubLink merger trees \citep{RodriguezGomez2015MNRAS}. For each halo, we follow the main progenitor branch of its central galaxy and record the halo mass $M_{\rm 200c}$ at each snapshot, which gives the mass accretion history $M_{\rm 200c}(z)$. Limited by the friends-of-friends (FoF) algorithm, the subhalo that SubLink identifies as the main progenitor is in some snapshots a satellite within a larger group rather than the central galaxy of its own group. At such snapshots, the recorded halo mass belongs to that larger group, not to the progenitor of our halo. We therefore discard these snapshots from the mass accretion history. We then require the remaining history to extend to $z=3$ and exclude the 8 halos that fail this requirement\footnote{The eight excluded halos differ from the seven removed in \citetalias{Xu2026arXiv}. Here we use the {\tt SubLink\_gal} merger tree, which is based on the baryonic component, whereas \citetalias{Xu2026arXiv} used the default {\tt SubLink} tree, which is based on dark matter. This choice does not influence our main results.}.

    From the mass accretion history, we then calculate the accretion rate of each halo at each snapshot. Unlike the backward-looking rate used by \citet{Diemer2017ApJS} and \citetalias{Xu2026arXiv}, we average the halo mass growth over the \emph{following} half dynamical time:
    \begin{equation}
        \Gamma_{\rm dyn+}(t)=\frac{\ln[M(t)]-\ln[M(t+t_{\rm dyn}/2)]}{\ln[a(t)]-\ln[a(t+t_{\rm dyn}/2)]},
    \end{equation}
    where $M=M_{\rm 200c}$, $a$ is the scale factor, and $t_{\rm dyn}=2R_{\rm 200c}/V_{\rm 200c}$ is the dynamical time. 

% -------------------------------------------------------------------------------------------- %
\subsection{Fossil Correlation \& Rank Retention Fraction}
    \label{ssec:assess}

    To quantify how the assembly history of massive halos is encoded in present-day galaxy properties, we use two diagnostics that connect present-day galaxy properties to their past halos, as illustrated in Fig.~\ref{fig:illustration}.

    \paragraph{Fossil Correlation $\tilde{\rho}(z)$} We quantify how well present-day galaxy properties preserve information about past halo assembly by computing the Pearson correlation coefficient\footnote{The results are consistent when we use the Spearman coefficient.} $\tilde{\rho}(z)$ between the logarithm of a present-day ($z=0.4$) observable (e.g., central stellar mass or satellite richness) and the logarithm of a property of its main progenitor halo (e.g., halo mass or historic richness) at each earlier redshift, denoted as $\tilde{\rho}_{M_{\rm halo}}(z)$ or $\tilde{\rho}_{\lambda}(z)$. 

    We refer to the relation between a present-day stellar mass and the progenitor halo mass at an earlier epoch as the \emph{fossil SHMR}, and $\tilde{\rho}_{M_{\rm halo}}(z)$ is its correlation coefficient. It is not the stellar-to-halo mass relation of the halo population at that epoch; it describes how much information about the earlier halo mass a present-day galaxy retains. The evolution of $\tilde{\rho}(z)$ with redshift reveals which cosmic epochs are most informative for reconstructing halo assembly from present-day observations.

    In addition, we ask whether the growth of a halo after redshift $z$ carries information about a present-day observable beyond the assembled \mhalo{} at that moment. We compute the partial fossil correlation $\tilde{\rho}_{\Gamma_{\rm dyn+}|M_{\rm halo}}(z)$: the Spearman correlation between the present-day observable and the forward accretion rate $\Gamma_{\rm dyn+}(z)$ (\S\ref{ssec:hist}) after the dependence of both on $M_{\rm halo}(z)$ is removed. A positive value means that, among halos with the same mass at $z$, those that grow faster during the following half dynamical time have larger present-day values of the observable. A value consistent with zero means that this growth adds no further information. The $\tilde{\rho}_{\Gamma_{\rm dyn+}|M_{\rm halo}}(z)$ allow us to associate the galaxy assembly with specific epochs of accretion \footnote{For halo mass defined with spherical overdensity, such as $M_{\rm 200c}$, $t_{\rm dyn}=0.2/H(z)$ is the same for all halos at a given redshift, so $\Gamma_{\rm dyn+}$ is a linear function of $\ln M(t+t_{\rm dyn}/2)-\ln M(t)$. For Pearson coefficients, the partial correlation is then fully determined by three ordinary correlations, $(\rho_{X2}-\rho_{X1}\rho_{12})/\sqrt{(1-\rho_{X1}^2)(1-\rho_{12}^2)}$, where $\rho_{X1}$ and $\rho_{X2}$ are the fossil correlations at $t$ and $t+t_{\rm dyn}/2$, and $\rho_{12}$ is the correlation between the two halo masses. The Spearman version used here differs from the Pearson version by $\lesssim0.05$ (rms).} Uncertainties on all correlation coefficients are estimated via bootstrap resampling.

    \paragraph{Rank Retention Fraction $f_{\rm ret}$} In addition, we want to quantify the current observable's ability to select the most massive halos at earlier epochs. For this purpose, we define a Top-$N$ sample using a present-day observable (e.g., the most massive 100 galaxies using $M_{\star,30}$) and calculate what fraction of this sample is \emph{retained} in the most massive $N$ halos at different epochs using their main progenitors' mass assembly history (see Fig.~\ref{fig:illustration}). As in \cite{Onions2025MNRAS}, the rank retention fraction $f_{\rm ret}$ is the fraction of present-day Top-$N$ halos whose progenitors exceed an evolving halo-mass threshold set by the $N$ value. Relative to fossil correlation, what the rank retention fraction adds is the consequence for sample selection: it measures directly how many of the halos selected today by a given observable were among the most massive halos at an earlier epoch.

    By construction, $f_{\rm ret}=1$ at $z=0.4$ for a selection by halo mass itself. At earlier epochs it falls below unity both because the halo-mass ranking is reshuffled and because some of the most massive progenitors later merge into other halos or end up below our mass limit, so $f_{\rm ret}$ measures the progenitor bias of each selection directly. We estimate its uncertainty by jackknife resampling of 12 subvolumes of the simulation box.

% -------------------------------------------------------------------------------------------- %
% Results
% -------------------------------------------------------------------------------------------- %
\section{Results} 
    \label{sec:results}

% -------------------------------------------------------------------------------------------- %
\subsection{Fossil Correlation: Stellar Radial Profile as Halo Assembly Record}
    \label{ssec:pastcorr}

\begin{figure*}
    \centering
    \includegraphics[width=1.0\linewidth]{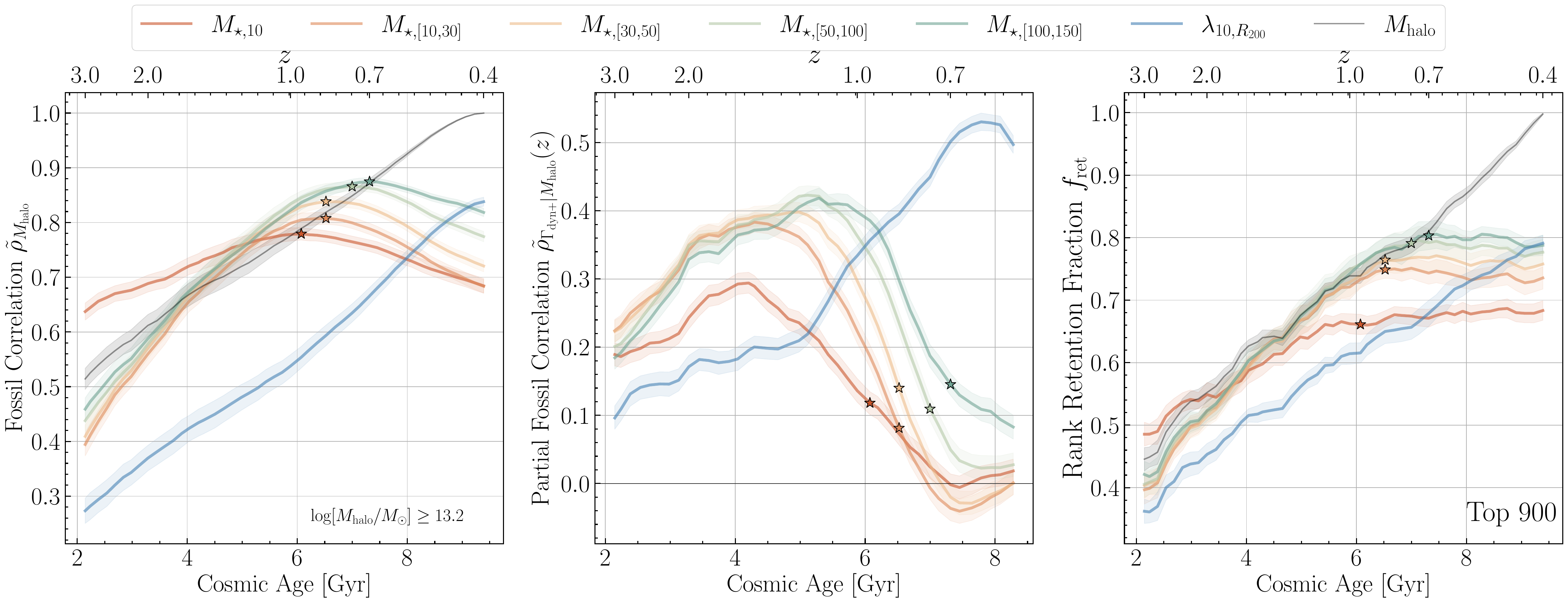}
    \caption{
        \textbf{Left panel:} Fossil correlation between present-day central stellar mass in different radial regions (color-coded) and past halo mass $M_{\rm halo}(z)$. The blue line shows richness $\lambda_{10, R_{200}}$, and the gray line shows $M_{\rm halo}$ at $z=0.4$. All radial components exhibit a characteristic non-monotonic evolution with a peak at intermediate redshift, but the peak shifts to later epochs for larger radii. \textbf{Middle panel:} Partial fossil correlation between present-day stellar mass and the forward accretion rate $\Gamma_{\rm dyn+}(z)$ at fixed halo mass at the same redshift (\S\ref{ssec:assess}). Stars mark the fossil correlation peaks from the left panel. The curves end at $z\approx0.55$ because $\Gamma_{\rm dyn+}$ requires the following half dynamical time. The radial dependence mirrors the left panel: outer regions retain positive partial correlations to later times, while the partial correlation for richness rises toward low redshift.
        \textbf{Right panel:} Rank retention fraction for the same radial selections (Top-$N=900$, selected from the full sample). The innermost stellar mass ($M_{\star,10~{\rm kpc}}$) exhibits the highest rank retention at $z\gtrsim2$, staying nearly constant at low redshift and declining at $z\gtrsim1.1$, whereas outer components exhibit the same two-phase behavior, but the transition into the declining phase shifts to progressively lower redshifts (later epochs) for larger radii. Stars mark the fossil correlation peaks from the left panel. Collectively, these panels demonstrate that the stellar mass profile preserves a radially stratified memory of halo assembly: the inner regions best encode the halo mass at earlier epochs, and the outer regions at later epochs. The shaded regions indicate the error bars, estimated from bootstrapping in the left and middle panels and from jackknife resampling in the right panel.
    }
    \label{fig:profile}
\end{figure*}

    To investigate how halo assembly history is encoded in the stellar distribution, we calculate the fossil correlation between present-day galaxy properties and progenitor halo masses at higher redshifts for the $M_{\rm halo}\geq10^{13.2}\, M_{\odot}$ sample ($N=2035$; Fig.~\ref{fig:profile}). 

    While the richness correlation weakens monotonically toward higher redshift, all stellar components of the central galaxy show non-monotonic correlations that peak at intermediate redshifts. This indicates a duality between the central galaxy and its satellites in tracing halo assembly: the satellites trace recent accretion, while the central galaxy serves as a fossil record of earlier halo assembly, consistent with the findings of \citetalias{Xu2026arXiv}.

    The redshift of these peaks depends on radius: the peak moves from $z_{\rm peak}\approx1.0$ for $M_{\star,10}$ to $z_{\rm peak}\approx0.7$ for $M_{\star,[100,150]}$, a shift much larger than the bootstrap uncertainties of the individual peaks ($\lesssim0.05$). 
    These peak positions are stable against changes in the threshold on the present-day halo mass between $10^{13.0}$ and $10^{14.0}\, M_{\odot}$, except for $M_{\star,10}$ in the two smallest, most massive samples (Appendix~\ref{app:massdep}).

    If a radial component recorded the halo mass at a single epoch, its fossil correlation at other epochs would follow the correlation between the progenitor halo masses at the two epochs, which declines smoothly away from that epoch (Appendix~\ref{app:single}). The annular components follow this prediction to within 0.05 between $z_{\rm peak}$ and $z\approx2$, whereas $M_{\star,10}$ lies well above it at high redshift, which indicates additional sensitivity to the earliest halo assembly epochs. Each radial component thus retains a record of the cumulative halo mass at a distinct epoch: \emph{the stellar mass profile constitutes a stratified fossil record of the halo assembly history.}
    
    Single-epoch recording makes a testable prediction. A component's correlation with the halo mass at its recording epoch is immune to dilution by later accretion, unlike the correlation with the present-day halo mass. The component can therefore correlate more strongly with the past halo mass than the present-day halo mass, provided it tracks the recording-epoch mass tightly enough. This is indeed the case: between their peaks and $z\approx1.5$, the annular components correlate with the past halo mass more strongly than the present-day halo mass itself does (gray curve in Fig.~\ref{fig:profile}), and $M_{\star,10}$ does so at all $z\gtrsim1$.  The present-day halo mass is a diluted record of the early halo assembly, while the stellar profile retains the undiluted version.

    The fossil correlation, however, is built on the cumulative mass $M_{\rm halo}(z)$, which mixes contributions from all epochs earlier than $z$. To measure how halo growth after each epoch contributes to each component, we use the partial fossil correlation $\tilde{\rho}_{\Gamma_{\rm dyn+}|M_{\rm halo}}(z)$ (middle panel of Fig.~\ref{fig:profile}).

    These partial correlations follow the same radial order as the fossil correlations. For every central stellar component, halo growth at $z\sim1-2$  carries additional information about the present-day stellar mass, and this information then declines toward lower redshift. The decline occurs later at larger radii: the partial correlation of the inner components declines earlier and vanishes by $z\approx0.8$, whereas that of the outer components declines progressively later, and the outermost annulus retains a positive partial correlation down to the last measurable epoch, $z\approx0.55$. These trends are nearly independent of the halo-mass threshold.

    By the fossil correlation peak of each central stellar component, the partial correlation has already declined to a low level ($\lesssim0.1$), and it vanishes at lower redshift. The fossil correlation peak therefore marks the end of the period during which halo growth is reflected in that component. Halo growth stops adding information about the central galaxy progressively later at larger radii. In this picture, growth at $z\sim1.5$--$2$ is reflected in all radial components, growth at $z\sim1$ mainly in the components beyond $\approx50$~kpc, and later growth mainly in the surviving satellite population.

    The innermost component is an exception. $M_{\star,10}$ maintains a strong fossil correlation with the progenitor halo mass out to $z\sim3$, but its partial correlation is the lowest of all components at $z\gtrsim1$ and declines earliest, although it approaches zero only slowly. The present-day $M_{\star,10}$ therefore tracks more closely the mass the halo had already assembled at early times than its subsequent growth, and retains the strongest memory of the earliest halo assembly epochs.
    
% -------------------------------------------------------------------------------------------- %
\subsection{Rank Retention Fraction: Progenitor Selection Bias of Massive Halos}
    \label{ssec:pastfrac}

    The rank retention fraction translates these correlations into the progenitor bias of samples selected by present-day observables. The rank retention fraction for the central stellar components shows a two-phase lookback evolution (Fig. \ref{fig:profile}): at low redshift it remains relatively high but begins to decline at some threshold redshift. In contrast, the fractions for halo mass and richness decrease across all redshifts. Beyond $z\approx0.7$, the fractions for all stellar components exceed that for richness, and at $z\gtrsim1$ those of the annular components are within 0.04 of the fraction for the present-day halo mass itself. This evolution also depends on radius: at larger radii, the plateau value is higher, and the transition between the two phases occurs at lower redshift.

    Richness counts the surviving satellites, which are continuously replenished by ongoing accretion, so it preferentially selects halos with active late-time assembly and suffers strong progenitor bias at high redshift. In contrast, the central stellar mass is most closely tied to the halo mass at earlier epochs and is less sensitive to later halo growth (\S\ref{ssec:pastcorr}). It therefore tends to select halos that were already massive at those epochs, maintaining high progenitor fidelity over a wider lookback baseline.

    We define the transition as the break point of a continuous two-segment linear fit to $f_{\rm ret}$ as a function of cosmic time. The transitions move from a cosmic age of 5.5~Gyr for $M_{\star,10}$ to 6.4~Gyr for $M_{\star,[100,150]}$. They follow the same radial ordering as the fossil correlation peaks but occur 0.6--0.9~Gyr earlier. Some correspondence between the two diagnostics is expected on statistical grounds: for a fixed sample size, the overlap between two Top-$N$ selections is largely set by the correlation between the two ranking quantities. 

% -------------------------------------------------------------------------------------------- %
% Discussions
% -------------------------------------------------------------------------------------------- %

\section{Discussion} 
    \label{sec:discussions}

\subsection{In Situ and Ex Situ Contributions to the Stratified Fossil Record}
\label{ssec:insexs}

    \begin{figure*}
        \centering
        \includegraphics[width=1\linewidth]{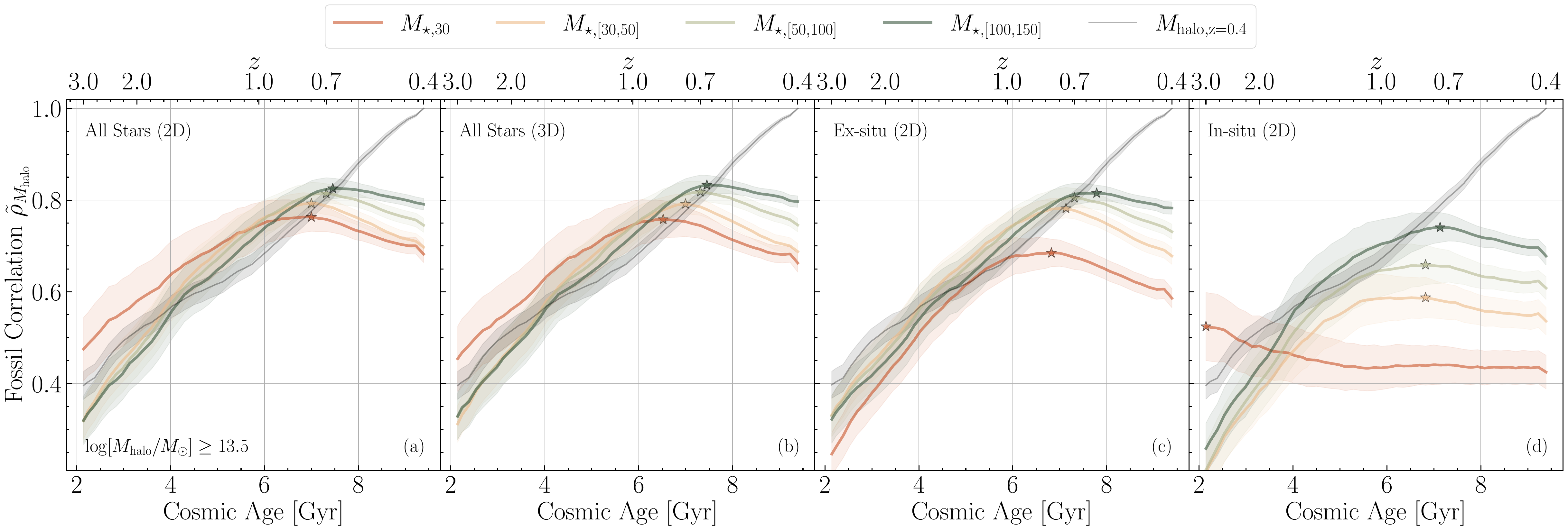}
        \caption{
            Fossil correlation between the past halo mass $M_{\rm halo}(z)$ and various stellar mass measurements for the massive-halo sample with $\log[M_{\rm halo}/M_\odot]\geq 13.5$ at $z=0.4$. Colored curves show the fossil correlation for the stellar mass within different regions; the black curve shows the fossil correlation for $M_{\rm halo}$ at $z=0.4$. Shaded bands indicate the $1\sigma$ error from bootstrapping. \textbf{Panel (a)}: Fossil correlation for the central stellar mass measured in 2D elliptical apertures, using the same aperture definition as Fig.~\ref{fig:profile} but with a coarser radial binning (innermost aperture of $30\,$kpc). \textbf{Panel (b)}: Same as panel (a), but with the stellar mass measured in 3D shells; the resulting curves are consistent with those in panel (a). \textbf{Panel (c)}: Fossil correlation for the \exsitu{} stellar mass in 2D apertures; the curves are also stratified by aperture radius, with outer apertures showing stronger correlations. \textbf{Panel (d)}: Fossil correlation for the \insitu{} stellar mass in 2D apertures; outer apertures show weaker, broader peaks that shift to later epochs with radius, while the innermost aperture rises monotonically toward higher redshifts. 
        }
        \label{fig:components}
    \end{figure*}
    
    In the two-phase picture (\S\ref{sec:intro}; \citealt{Naab2009ApJl}; \citealt{Oser2010ApJ}), \insitu{} star formation builds the compact core early, while accreted satellites progressively build the outer stellar halo. We first verify that the stratified fossil correlation is not a projection artifact, then use the \insitu{}/\exsitu{} decomposition to identify the origin of the stratified fossil record. We use the $\log[M_{\rm halo}/M_{\odot}]\geq13.5$ subsample ($N=$910), for which the \insitu{} and \exsitu{} mass decomposition is available (\citealt{RodriguezGomez2016MNRAS}; \citealt{Zhou2025}). Because the fossil-correlation peaks are insensitive to the halo-mass threshold (\S\ref{ssec:pastcorr}), this subsample can be compared directly with the fiducial sample of Fig.~\ref{fig:profile}. In this subsample, the median \exsitu{} fraction rises from $\sim$65\% within 30~kpc to $\sim$85\% in the [100,150]~kpc annulus. 

    The fossil correlations of the stellar mass measured in 2D elliptical apertures (panel a of Fig.~\ref{fig:components}) and in 3D ellipsoidal shells (panel b)\footnote{To ensure a fair comparison between the 2D and 3D masses, the 3D shells are placed at correspondingly larger radii; e.g., the [50,100]~kpc 2D aperture corresponds to a [79,158]~kpc 3D shell \citep{Zhou2025}.} are nearly identical, so the stratified fossil record reflects an intrinsic property of the central galaxy in \tng{}.

    For the \exsitu{} mass (panel c), the curves remain radially stratified and retain the peak migration seen for the total stellar mass. However, the enhanced early-epoch correlation of the innermost component seen in panel (a) disappears: at $z\approx3$, the correlation of $M_{\star,30}$ drops from 0.47 for all stars to 0.24 for the \exsitu{} mass, whereas it reaches 0.52 for the \insitu{} mass.
    For the \insitu{} mass (panel d), the outer components also show peaked curves whose peaks shift to later epochs with radius, but these peaks are weaker and broader than those of the \exsitu{} mass and lie at slightly earlier epochs, while the correlation for the innermost \insitu{} mass increases monotonically toward higher redshift. This disappearance of the early-epoch enhancement in the correlation of the innermost \exsitu{} component, together with the monotonic early-epoch rise of the \insitu{} component, identifies the \insitu{} channel as the carrier of the early-epoch memory inferred in \S\ref{ssec:pastcorr}. 

    These results indicate that the \exsitu{} stars, which drive the substantial size growth of massive galaxies, carry the radial shift of the fossil-correlation peaks, whereas the \insitu{} stars carry the early-epoch memory of the innermost region, consistent with the two-phase formation picture (\citealt{Oser2010ApJ}; \citealt{vanDokkum2010ApJ}; \citealt{RodriguezGomez2016MNRAS}). Because the \exsitu{} fraction increases with radius, the two channels together produce the radial stratification of the fossil record. If the outer stellar halo is indeed built from accreted satellite debris, it should retain a memory not only of the halo mass assembly but also of the past satellite population itself. We test this directly in the next section.

\subsection{The Stellar Halo Traces Historic Richness}
\label{ssec:histrich}

\begin{figure}
        \centering
        \includegraphics[width=0.9\linewidth]{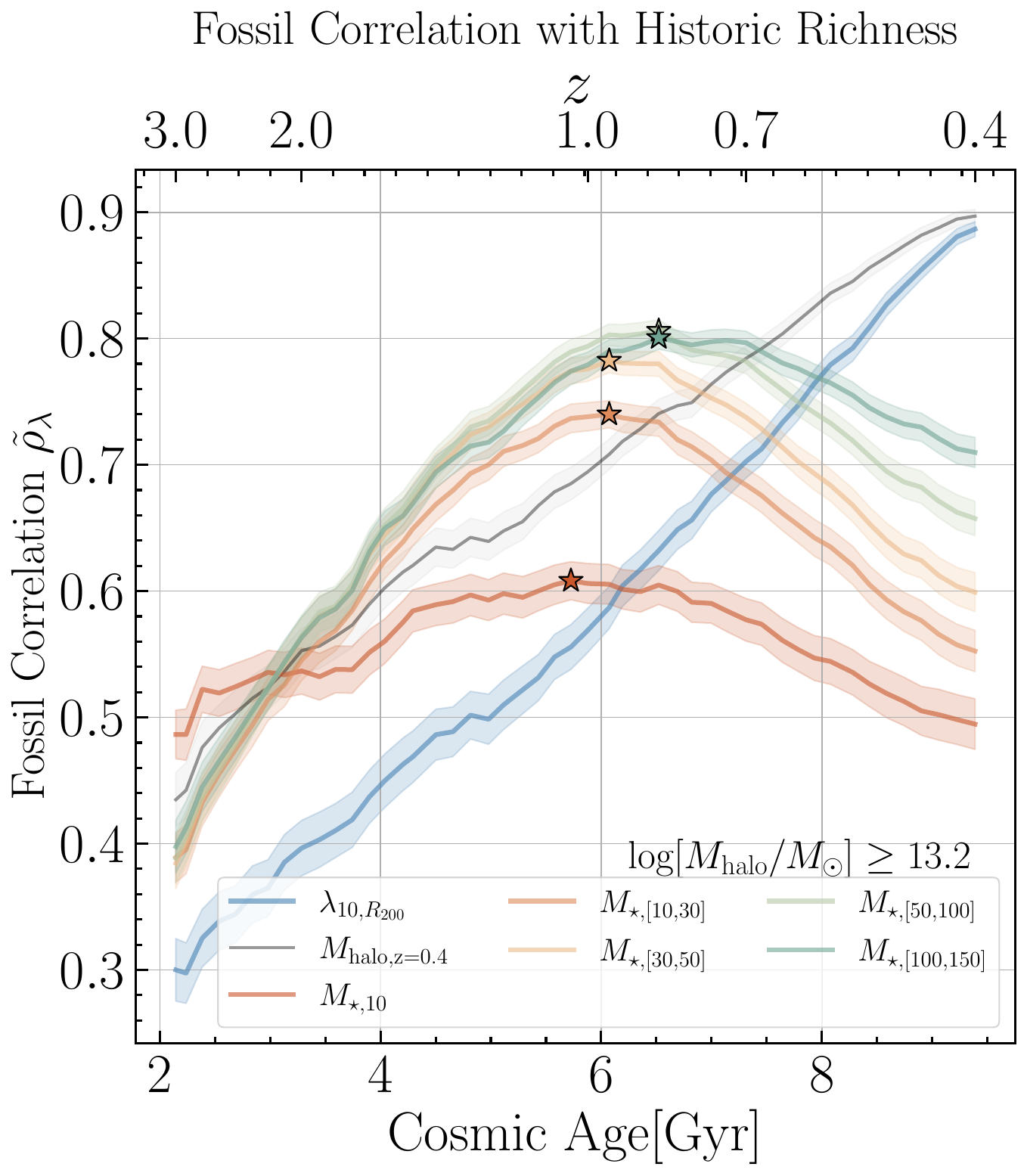}
        \caption{
            Fossil correlation between present-day galaxy properties and past satellite richness ($\lambda_{9, R_{200}}$) as a function of cosmic age. All radial components peak near $z\approx1$, with a smaller shift with radius than in Fig.~\ref{fig:profile}, while the halo mass and richness correlations increase monotonically toward the present. Importantly, the correlation is stronger in the outer regions than in the inner regions. Furthermore, stellar mass in the innermost region ($10 \text{ kpc}$) shows a notably weaker correlation than its correlation with past halo mass shown in Fig. \ref{fig:profile}, suggesting that inner-region stellar mass growth is insensitive to richness evolution.
        }
        \label{fig:pastrich}
    \end{figure}

    The \exsitu{} origin of the outer stellar halo makes a direct prediction: the stellar halo, built from tidally stripped and disrupted satellites, should also encode information about the past satellite population. We test this by computing the fossil correlation $\tilde{\rho}_{\lambda}$ between the present-day galaxy properties and the historic richness, defined as the number of satellites at earlier redshifts (Fig. \ref{fig:pastrich}). These fossil correlations also peak. In contrast to $\tilde{\rho}_{M_{\rm halo}}$, their amplitudes show a striking radial dependence: the inner components couple weakly, whereas the outer stellar halo couples strongly.
    
    The inner stellar mass ($M_{\star,10}$) couples much more weakly with historic richness than with past halo mass. This reflects the different assembly channels traced by the two quantities: the halo mass includes the smoothly accreted component that fuels star formation, whereas the richness counts the surviving satellite population. The inner region, where the \insitu{} fraction is highest, is expected to follow the former but not the latter. 
    
    In contrast, the outer stellar halo correlates strongly with historic richness: at its peak, at a cosmic age of 6.5~Gyr, the correlation reaches 0.80, compared with 0.60 for present-day richness and 0.74 for present-day halo mass at the same epoch. Physically, the present-day richness counts the surviving, continuously replenished satellite population, while the outer stellar halo preserves the integrated record of disrupted satellites. This enhanced coupling therefore shows that the outer stellar halo retains information about the satellite population at intermediate redshift. The correlations with historic richness also peak later at larger radii, but by less: from $M_{\star,[10,30]}$ to $M_{\star,[100,150]}$, their peaks move from a cosmic age of 6.1 to 6.5~Gyr, compared with 6.4 to 7.3~Gyr for $\tilde{\rho}_{M_{\rm halo}}$ (Figs.~\ref{fig:profile} and \ref{fig:pastrich}), about half the shift. Unlike the cumulative halo mass, the historic richness counts only the satellites that survive at each epoch: at earlier epochs many future contributors to the stellar halo have not yet arrived, and at later epochs the earliest contributors have already been disrupted. The correlation thus peaks when the surviving satellites best overlap the contributor population. The tendency for later-accreted contributors to deposit at larger radii shifts this epoch to later times for the outer components, but less than for $\tilde{\rho}_{M_{\rm halo}}$, because the census itself carries no radial information: the epoch of maximal overlap is largely set by the satellite population as a whole.
    
     The stellar halo thus offers a diagnostic of past satellite accretion that the present satellite census alone cannot provide.

% -------------------------------------------------------------------------------------------- %
% Summary and Conclusions
% -------------------------------------------------------------------------------------------- %

\section{Summary and Conclusions} 
    \label{sec:summary}

   Using massive halos at $z=0.4$ in {\tt TNG300-1}, we measured how the stellar mass profiles of their central galaxies encode past halo assembly, with two diagnostics: the fossil correlation $\tilde{\rho}(z)$ and the rank retention fraction $f_{\rm ret}(z)$. Our main findings are:

   \begin{itemize}
        \item The stellar profile of massive central galaxies (hosted by halos with $M_{\rm 200c}\gtrsim10^{13}~M_\odot$) preserves a radially stratified fossil record of halo assembly history. The fossil correlation peaks shift from $z\approx1.0$ for $M_{\star,10}$ to $z\approx0.7$ for $M_{\star,[100,150]}$, consistent with the inside-out assembly picture. The peak redshifts are robust to the selection on present-day halo mass, whereas the correlation amplitudes are not.
        
        \item Each fossil correlation peak marks the end of the epoch when the corresponding component was most sensitive to halo accretion. This sensitivity ends progressively later at larger radii. The innermost component tracks the halo mass assembled at earlier epochs.

        \item The annular stellar masses of the central galaxy beyond 10 kpc are less biased tracers of the most massive halos at $z>0.7$ than the current richness. At $z\gtrsim1$, they recover these halos as reliably as the present-day halo mass.

        \item The radial stratification is an intrinsic 3D property of massive central galaxies not a projection artifact.
               
        \item The \exsitu{} stars carry the radial stratification: the \exsitu{} stellar mass retains the stratified, peak-migrating correlations, whereas the \insitu{} mass shows weaker and broader peaks that also shift with radius. The \insitu{} mass in the innermost region retains the memory of the earliest epochs. 
        
        \item The outer stellar halo traces the satellite population at intermediate redshift better than either present-day richness or present-day halo mass.    
   \end{itemize}

   The fossil correlation and rank retention fraction provide a quantitative basis for forward models that map a halo's mass accretion history onto the central stellar mass profile and associated satellite population.

% -------------------------------------------------------------------------------------------- %
% Acknowledgements
% -------------------------------------------------------------------------------------------- %

\section*{Acknowledgment}

  % SH's funding
  SH acknowledges support from the National Natural Science Foundation of China Grant No. 12273015 \& No. 12433003.
  
  % AL & BD's funding
  AL acknowledges the use of the lux supercomputer at UC Santa Cruz, funded by NSF MRI grant AST
  1828315. AL is supported by the U.S. Department of Energy, Office of Science, Office of High
  Energy Physics under Award Number DE-SC0019301. AL acknowledges support from the David and
  Lucille Packard Foundation, and from the Alfred P. Sloan Foundation.
  BD gratefully acknowledges support from the Sloan Foundation.

  % IllustrisTNG
  The IllustrisTNG simulations were undertaken with compute time awarded by the Gauss Centre for Supercomputing (GCS) under GCS Large-Scale Projects GCS-ILLU and GCS-DWAR on the GCS share of the supercomputer Hazel Hen at the High Performance Computing Center Stuttgart (HLRS), as well as on the machines of the Max Planck Computing and Data Facility (MPCDF) in Garching, Germany. We thank the IllustrisTNG collaboration for making the simulation data publicly available \citep{Nelson2019ComAC}.

% -------------------------------------------------------------------------------------------- %
% Data Availability
% -------------------------------------------------------------------------------------------- %
\section*{Data Availability}

  The IllustrisTNG simulation data used in this work are publicly available at \url{https://www.tng-project.org/} \citep{Nelson2019ComAC}. The derived catalogs used in this paper, including the aperture and annular stellar masses of the central galaxies, the present-day and historic richness, the main-progenitor mass accretion histories and accretion rates, and the fossil correlations and rank retention fractions shown in the figures, will be released together with the figure scripts in a GitHub repository (\url{https://github.com/Xuchuyi/radial_fossil}).
  % and archived on Zenodo (\dataset[doi:10.5281/zenodo.XXXXXXX]{https://doi.org/10.5281/zenodo.XXXXXXX}; \textcolor{red}{placeholder}).

% Keep track of all the software used in this work
\software{
    \href{http://www.numpy.org}{\code{NumPy}} \citep{Numpy},
    \href{https://www.scipy.org}{\code{SciPy}} \citep{scipy, Virtanen2020NatMe},
    \href{https://www.astropy.org/}{\code{Astropy}} \citep{astropy, Astropy2018AJ, Astropy2022ApJ},
    \href{https://photutils.readthedocs.io}{\code{Photutils}} \citep{Photutils},
    \href{https://matplotlib.org}{\code{Matplotlib}} \citep{matplotlib}
}

% -------------------------------------------------------------------------------------------- %
% Appendix
% -------------------------------------------------------------------------------------------- %
\appendix
\twocolumngrid
\renewcommand{\thefigure}{\Alph{section}\arabic{figure}}
\section{Mass dependence and peak robustness of the fossil correlation}
\label{app:massdep}

\setcounter{figure}{0}

\begin{figure*}
        \centering
        \includegraphics[width=\linewidth]{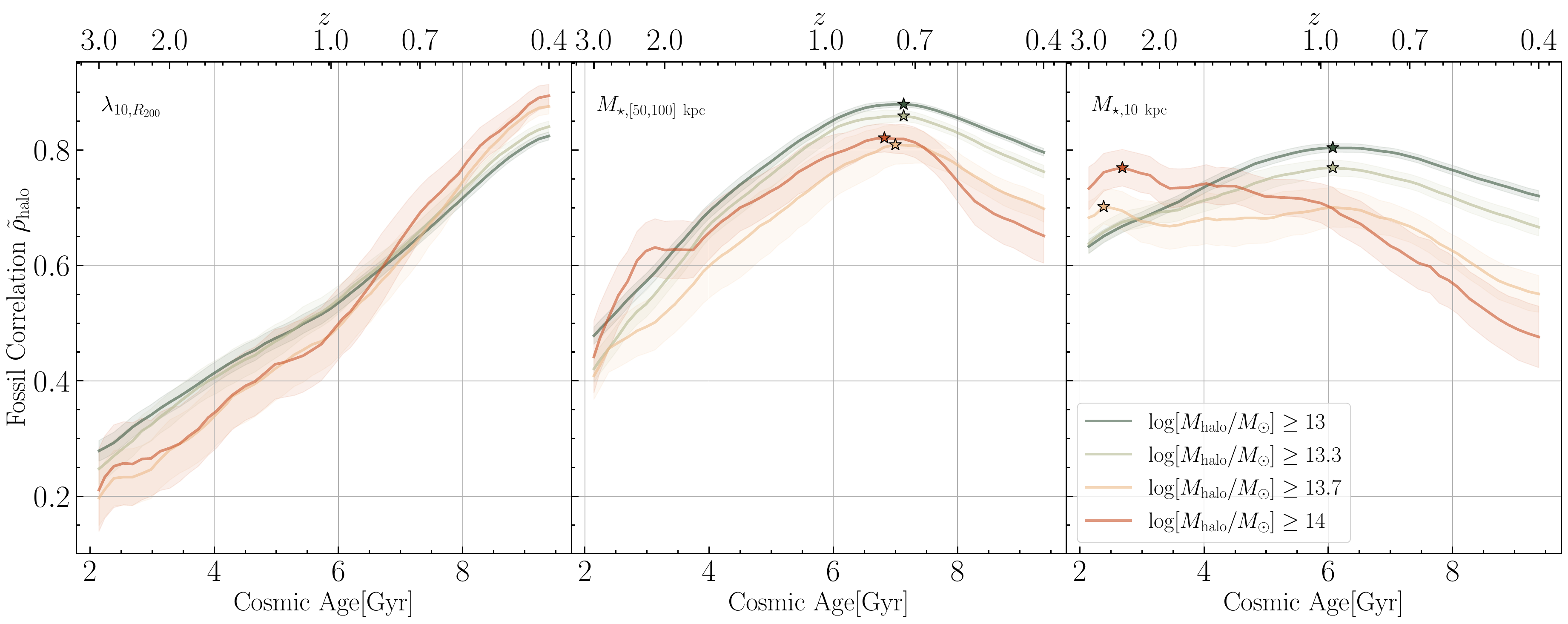}
        \caption{
            Halo mass dependence of present-day galaxy property correlation with past halo mass across cosmic time. From left to right, panels show the correlation $\rho$ for richness $\lambda_{10,R_{200}}$, outer stellar mass $M_{\star,[50,100]\text{ kpc}}$, and inner stellar mass $M_{\star,10\text{ kpc}}$, categorized by different minimum halo mass thresholds. Colors indicate different minimum halo mass thresholds. For richness, all mass scales increase monotonically, but more massive halos show a stronger correlation with their past mass at low redshift ($z\lesssim0.7$) than lower-mass systems. For outer and inner stellar mass, the peaks for different thresholds occur at similar redshifts (except the two highest halo mass samples for innermost region). Lower-mass halos show higher peaks but decline more rapidly toward high redshift, while massive halos maintain more persistent but weaker correlations.}
        \label{fig:diffmass}
    \end{figure*}

    Because halos of different masses have systematically different assembly histories (e.g., \citealt{Wechsler2002ApJ}; \citealt{McBride2009MNRAS}; \citealt{Zhao2009ApJ}), we examine whether the fossil correlations identified in \S\ref{ssec:pastcorr} persist across the mass range. We compare four samples with minimum halo masses of $\log[M_{\rm halo}/M_\odot]=13$, 13.3, 13.7, and 14, and compute the fossil correlations $\tilde{\rho}_{M_{\rm halo}}$ for richness ($\lambda_{10, R_{200}}$) and for the central stellar masses $M_{\star,[50,100]~{\rm kpc}}$ and $M_{\star,30~{\rm kpc}}$ (Fig.~\ref{fig:diffmass}).

    The correlation amplitudes and detailed curve shapes vary with halo mass. Samples with lower mass thresholds reach higher peak correlations for central stellar mass. Richness shows the opposite trend at low redshift. Its correlation with past halo mass is stronger in the more massive samples. For the stellar components, the peaks of the different mass samples occur at similar epochs in Fig. \ref{fig:diffmass}. This agreement suggests that the peak positions are less sensitive to halo mass than the correlation amplitudes.

    To test the robustness of the peak positions directly, we measure $z_{\rm peak}$ for all five central stellar components used in Fig. \ref{fig:profile}. We use six samples with minimum present-day halo masses of $\log[M_{\rm halo}/M_{\odot}]=13.0$, 13.2, 13.3, 13.5, 13.7, and 14.0. For each sample and each component, we first identify the snapshot with the highest fossil correlation. We then fit a quadratic function to $\tilde{\rho}_{M_{\rm halo}}$ as a function of cosmic age within $\pm 0.5$ Gyr of this maximum. We adopt the vertex of the quadratic fit as the peak cosmic age and then convert this age to $z_{\rm peak}$. We estimate the peak uncertainties with 500 bootstrap realizations.

\begin{figure}[t]
        \centering
        \includegraphics[
            width=0.9\columnwidth
        ]{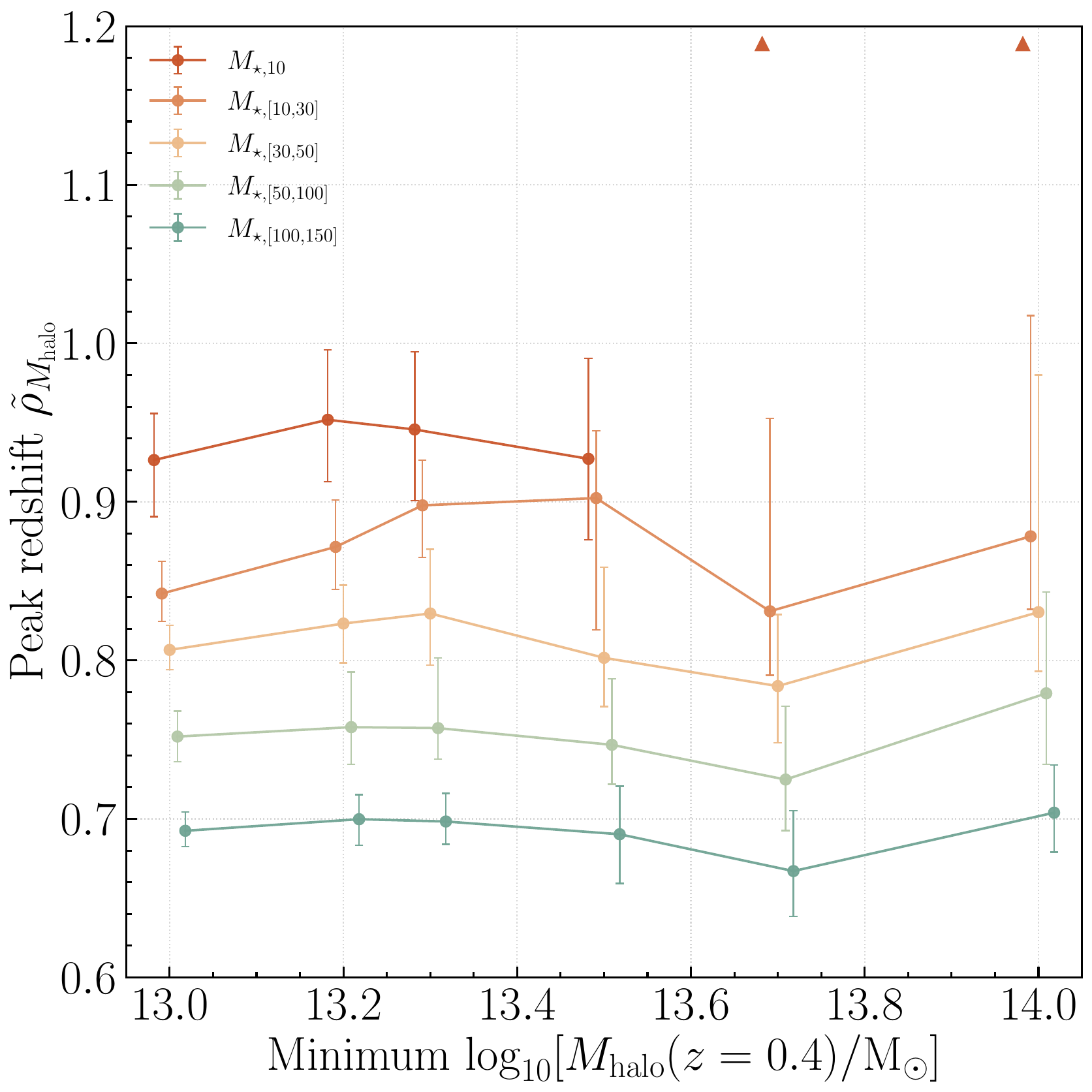}
        \caption{
            Peak redshifts of the fossil correlation
            $\tilde{\rho}_{M_{\rm halo}}$ as a function of the minimum present-day halo mass.
            Different colors show the five central stellar components used in Fig.~\ref{fig:profile}.
            We measure peak positions with local quadratic fits within $\pm 0.5$ Gyr of the maximum correlation. Error bars show the 16th--84th percentile ranges from 500 bootstrap realizations. We apply small horizontal offsets to the different components for clarity.
            Upward-pointing triangles at the upper plotting boundary denote $M_{\star,10}$ measurements whose fitted peaks lie above $z=1.2$. We omit their error bars because the corresponding correlation curves are broad and the peaks are poorly constrained.
            The outer components show weak variation with the mass threshold and preserve their radial ordering.
        }
        \label{fig:peak_mass}
    \end{figure}

    Figure~\ref{fig:peak_mass} shows the measured peak redshifts as a function of the minimum present-day halo mass. The four annular components show no systematic variation in $z_{\rm peak}$ across the halo mass thresholds. Their radial ordering also remains unchanged. The epochs encoded by these regions therefore do not depend on the choice of the present-day halo-mass threshold. This test directly supports the robustness of the radial stratification identified in \S\ref{ssec:pastcorr}.

    The innermost component, $M_{\star,10}$, is the only exception. Its fitted peak lies above $z=1.2$ for the two highest mass thresholds. The corresponding correlation curves are broad and nearly flat over a wide redshift range. These fitted values should therefore not be interpreted as precise peak measurements. The behavior of $M_{\star,10}$ at the highest mass thresholds may reflect both the different assembly histories of the most massive halos and the larger role of in situ star formation in their central regions. The fossil signal in this region may also be more sensitive to the details of the star formation and feedback models. Tests with other hydrodynamical simulations are needed to establish whether this behavior is generic.

    In summary, the amplitudes and detailed shapes of the fossil correlations depend on halo mass. However, the peak positions of the annular components show little variation with the present-day halo-mass threshold. Their radial ordering also remains stable across the full mass range. The radial stratification identified in \S\ref{ssec:pastcorr} is therefore a robust feature of the massive galaxy population.

\section{A Single-epoch Model for the Fossil Correlation}
\label{app:single}

    Here we derive the fossil correlation expected for a stellar component that records the halo mass at a single epoch, and compare it with the measured curves. Suppose that the present-day stellar mass of a radial component is determined by the progenitor halo mass at a single epoch $z_p$, $\log M_{\star}\simeq f(\log M_{\rm halo}(z_p))+\epsilon$, where the scatter $\epsilon$ is independent of the halo history. Its fossil correlation with the progenitor halo mass at any other epoch is then inherited from the correlation between the progenitor masses themselves, $\tilde{\rho}_{M_{\rm halo}}(z)\simeq A\, C_{\rm halo}(z_p,z)$, where $C_{\rm halo}(z_p,z)$ is the correlation between $\log M_{\rm halo}(z_p)$ and $\log M_{\rm halo}(z)$ across the sample and the constant $A\leq1$ absorbs the scatter $\epsilon$. Because the progenitor mass is cumulative, $C_{\rm halo}(z_p,z)$ declines smoothly away from $z_p$, so this model predicts a peak at $z_p$ with broad wings rather than a sharp cutoff.

    We test this prediction by dividing each measured $\tilde{\rho}_{M_{\rm halo}}(z)$ by its peak value and comparing it with $C_{\rm halo}(z_{\rm peak},z)$ computed from the same halos. For the annular components, the two agree to within 0.05 between $z_{\rm peak}$ and $z\approx2$. At later epochs, the normalized correlations lie slightly above the prediction, by up to 0.06 at $z=0.4$, so these components also retain some information about later halo growth; at $z\gtrsim2$ they fall below it, by 0.05--0.1. In contrast, the normalized correlation of $M_{\star,10}$ lies well above the prediction on both sides of its peak, by 0.2 at $z=3$.

% -------------------------------------------------------------------------------------------- %
% End of the Draft
% -------------------------------------------------------------------------------------------- %

\bibliography{histrich2}{}
\bibliographystyle{aasjournal}

%\listofchanges
\end{CJK*}
\end{document}